\documentstyle[twocolumn,prl,aps,graphicx]{revtex}
\begin{document}
\draft
\twocolumn[\hsize\textwidth\columnwidth\hsize\csname@twocolumnfalse\endcsname
\title{
       Evidence of short time dynamical correlations in simple liquids.
      }
\author{
        T.~Scopigno$^{1}$,
        G.~Ruocco$^{1}$,
        F.~Sette$^{2}$,
        G.~Viliani$^{3}$
        }
\address{
    $^{1}$Dipartimento di Fisica and INFM, Universit\'a di Roma ``La Sapienza'',
I-00185, Roma, Italy.\\
    $^{2}$European Synchrotron Radiation Facility, BP 220, F-38043 Grenoble
Cedex, France.\\
    $^{3}$Dipartimento di Fisica and INFM, Universit\'a di Trento, I-38100,
Povo, Italy.}
\date{\today}
\maketitle

\begin{abstract}
We report a molecular dynamics (MD) study of the collective dynamics of a
simple monatomic liquid -interacting through a two body potential that mimics
that of lithium- across the liquid-glass transition. In the glassy phase we
find evidences of a fast relaxation process similar to that recently found in
Lennard-Jones glasses. The origin of this process is ascribed to the
topological disorder, i.e. to the dephasing of the different momentum $Q$
Fourier components of the actual normal modes of vibration of the disordered
structure. More important, we find that the fast relaxation persists in the
liquid phase with almost no temperature dependence of its characteristic
parameters (strength and relaxation time). We conclude, therefore, that in the
liquid phase well above the melting point, at variance with the usual
assumption of {\it un-correlated} binary collisions, the short time particles
motion is strongly {\it correlated} and can be described via a normal mode
expansion of the atomic dynamics.
\end{abstract}
\pacs{PACS numbers: 63.50.+x, 64.70.Pf, 61.43.Fs, 62.60.+v}
]

\section{INTRODUCTION}
Many experimental investigations by Inelastic Neutron (INS) and X-ray (IXS)
Scattering techniques, as well as different theoretical and numerical studies
have been devoted since more than twenty years to the understanding of the
collective dynamics of simple liquids. In particular, the efforts have been
focused on those dynamical ranges that go beyond the two limiting cases of
single particle (high exchanged momentum, $Q$) and hydrodynamic (low $Q$)
regimes. The alkali metals are a typical class of systems considered as a
workbench to test the different theoretical approaches developed so far for
the dynamics of the liquid state. These systems are known to support
well-defined oscillatory modes for the density fluctuations (acoustic-like
modes) well outside the strict hydrodynamic region, down to wavelengths of a
few inter-particle distances.

From the experimental point of view, it is worth mentioning the pioneering
inelastic neutron scattering (INS) experiment by Copley and Rowe
\cite{rubidio2} in liquid rubidium, while more recently many experimental
efforts have been performed aiming to realize more and more accurate
experiments: INS investigations have been devoted to liquid cesium
\cite{cesio}, sodium \cite{sodio}, lithium \cite{verkerk}, potassium
\cite{potassio} and again rubidium \cite{rubidio}. Simultaneously many
theoretical and numerical studies have also been reported on the same systems
\cite{umb1,umb2,kamb,can,cas1,cas2,patacca,yulmetyev}. Among others, one
significant advantage of the numerical techniques with respect to INS is the
possibility to overcome the experimental restriction, i.~e. the limited
($Q-E$) accessible region and the simultaneous presence of coherent and
incoherent contributions to the INS signal. Moreover, numerical techniques
allow to access a wider set of correlation functions while the inelastic
scattering experiments basically probe the density-density correlation
function.

More recently, the advent of Inelastic X-ray Scattering (IXS) technique
\cite{noiV,noiM}, opened new possibilities for the experimental determination
of the dynamic structure factor in the nanometer wavevector region. In fact,
the use of X-rays often allows to extend the accessible exchanged energy region
to neutrons (an important merit especially in these systems characterized by a
high value of the sound velocity), and to overcome the presence of the
incoherent scattering when one is also interested in the collective motion. On
the other hand, the resolution demand is particularly severe in this case, and
the scattered intensity rapidly decreases on increasing the atomic number of
the sample. This complementarity of IXS and INS allowed to improve the
comprehension of the microscopic dynamics in alkali metals through a number of
experiments performed in the last decade
\cite{burk,sinn,pilgrim,PRL,euro,scop,PREal,PREna}.

A basic idea suggested by Molecular Dynamics (MD) simulations \cite{umbook},
and recently demonstrated by IXS in the case of lithium \cite{PRL,scop},
aluminum \cite{PREal}, and sodium \cite{PREna}, is the existence of two
different time-scales which drive the viscous decay of the density
fluctuations. These two time-scales are believed to reflect two different
phenomena. The slower process (whose characteristic time spans from few
picosecond to seconds depending on the temperature) is responsible for the
highly {\it correlated} atomic motion. Its current understanding is provided
by the Mode Coupling Theory (MCT) approach, which gives a satisfactory
description of a large number of experimental and numerical results obtained
in glass forming systems \cite{Goetze1}. This process is usually referred to
as the $\alpha$ (structural) relaxation process. The faster process
(characteristic time in the sub-picosecond time range) is less understood,
and, in the normal liquid phase, is usually described by {\it un-correlated}
binary collisions between particles \cite{sjogr}.

Recently, theoretical (hard spheres) \cite{Goetze2} and numerical
(Lennard-Jones) \cite{har} studies on the high frequency dynamics of model
monatomic {\it glasses} have shown that, even in a harmonic glass, it exists a
fast relaxation process. This process shows the typical phenomenology of
relaxations, as the sound velocity increases with wavevector $Q$, in spite of
the fact that all the diffusive degrees of freedom are frozen and the
anharmonicity is absent. This findings point towards the association of this
relaxation process to the topological disorder \cite{har}: i.~e. the dephasing
of the different oscillatory components present in the dynamics of the density
fluctuations at a given $Q$ value gives rise to the decay of the correlation
functions. If this interpretation is correct, this fast relaxation process
should be a general property of disordered systems and, in particular, it must
also exists in the liquid phase.

In this Paper, we investigate whether in a simple monatomic liquid it is
possible to associate the fast relaxation process to the topological disorder
and to improve our understanding of this phenomenon beyond the present belief
based on (almost) un-correlated binary collisions. This study compares the
dynamical properties of a monatomic system (lithium) in its liquid,
supercooled, glassy and crystalline phases obtained by Molecular Dynamics (MD)
simulations. The analysis of quantities such as the speed of sound, the
acoustic attenuation, and the related relaxation times indicates the existence
of the two-relaxation processes in all disordered phases. The slow process,
whose relaxation time largely increases approaching the structural arrest
(glass transition), effectively behaves as the $\alpha$ relaxation. Most
importantly, {\it we find that the fast process is always present and its
features do not change between the liquid and glass phases}. This implies the
importance of the topological disorder in this fast dynamics even in the
liquid phase. Moreover, it also indicates that even at the short timescale
considered the atomic motion is highly correlated and cannot be described by
uncorrelated binary collisions.

\section{MOLECULAR DYNAMICS SIMULATION}

We numerically investigated the molecular dynamics of 2000
lithium atoms interacting by the Price-Tosi pseudopotential. This
interaction potential has been shown to accurately reproduce both
the structural and the dynamical properties \cite{umb2}. Lithium
has been chosen due to the availability of high resolution
experimental data on the dynamic structure factor $S(Q,\omega)$
in the liquid phase, which show the existence of the two
relaxations processes \cite{scop}.

The molecular dynamics simulations have been performed in the microcanonical
ensamble, utilizing a Verlet algorithm with an integration time $dt=0.5$ fs.
The system has been equilibrated at several different temperatures, in a
square box with periodic boundary conditions. To make closer contact with the
IXS experiments, the numerical densities, i.e. the box dimensions, have been
adjusted at each temperature to the experimental values \cite{ose}, as far as
the mass of each atom, chosen as the isotopic mixture corresponding to the
natural abundance ($m=1.15221\times 10^{-26}$ kg). The energy conservation,
all over the explored temperature range, was better than few parts in
$10^{-4}$.

The melting temperature of the selected inter-atomic interaction potential is
$T_m=450$ K, in good agreement with the experimental melting point. Below this
temperature, the "MD liquid" can be supercooled avoiding the crystallization,
and it attains the structural arrest at $T_c \approx 260$ K, a value
extrapolated from a power law fit of the calculated diffusion coefficient.
This value of $T_c$ is consistent with that estimated from the depth of the
effective interaction potential well, $\epsilon$. In fact, using the relation
$T_c \approx 0.47 \epsilon$, valid for the Lennard Jones potential
\cite{ange}, we derive for the Price pseudopotential $T_c = 270$ K. Below
$T_c$ it is no longer possible to equilibrate the system, but we can obtain
the simulated "lithium glass" by rapid quenching down to few Kelvin. Further
heating of the system up to (but below that) $T_c$ allows to investigate the
effect of the anharmonicity in the glassy configuration. We studied the system
in the range from 5 K to 1100 K, a region much wider than that accessible in
the real system. Indeed, the low and high $T$ limits cannot be reached
experimentally because of the unavoidable crystallization below $T_m$ or of the
high chemical reactivity of the hot liquid. At any of the selected
temperatures (above $T_g$), we equilibrated the system for few hundreds
picoseconds and then we stored the configurations, ${\bf r}_i(t)$, in order to
calculate the dynamic structure factor, defined as:

\begin{equation}
S(Q,\omega)=(1/N) FT \sum_{i,j} \langle e^{-i\bf Q \cdot {\bf
r}_i(t)} e^{i\bf Q \cdot {\bf r}_j(0)}\rangle.
\end{equation}

For the sake of simplicity, we turned around the calculation of
the correlation by making use of the Wiener-Kintchine theorem,
that simplify the computation of the dynamic structure factor as:

\begin{equation}
S(Q,\omega)=\frac{1}{N}\left|\emph{FT}\sum_i\left\langle
e^{-i\textbf{Q}\cdot\textbf{R}_i(t)}\right\rangle\right|^{2}
\end{equation}

The numerical resolution $2\pi/\Delta t$, being $\Delta t$ the
acquisition length, was set to $1.3$ ps$^{-1}$, while the 'free
spectral range' (i.e. the accessible frequency window), $2 \pi /
\delta t$  where $\delta t$ is the time separation of the stored
configurations, was set to $120$ ps$^{-1}$. Each $S(Q,\omega)$ at
a fixed $Q$ value was obtained as an average over different
statistically independent runs (up to 52 for each temperature)
and $\vec{Q}$ orientations in the simulation box.

\section{RESULTS}

\begin{figure}[ftb]
\centering
\includegraphics[width=.47\textwidth]{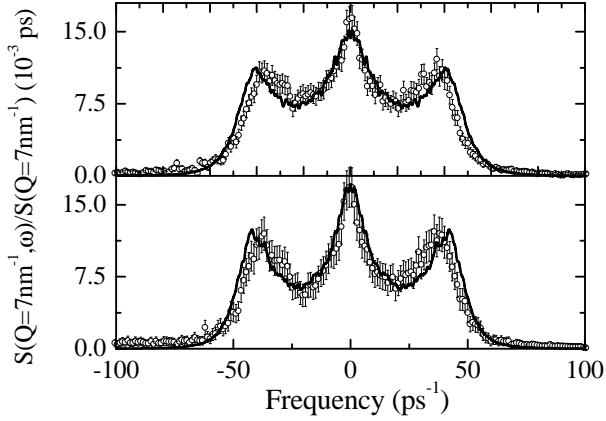}]
\vspace{-6.1cm} \caption{ Dynamic structure factors of lithium at
$Q=7$nm$^{-1}$ and $T=600$ K (upper panel) and $T=475$ K (lower panel) are
reported together with the experimental data (points with error bars)
\protect\cite{scop}. The resolutions (FWHM) are: $\delta E_{exp}\approx 4.5$
ps$^{-1}$ for the IXS data, $\delta E_{sim}=1.3$ ps$^{-1}$ for the
simulations. In the experimental data, the detailed balance factor has been
removed to be consistent with the classical character of the simulation. }
\label{panel_bis}
\end{figure}

In Fig.~\ref{panel_bis}, we report a comparison of the simulated dynamic
structure factor at $Q=7$ nm$^{-1}$ with the corresponding experimental result
for those temperatures where data are available (T=475 K and T=600 K).

The detailed balance factor has been removed from the experimental data through
the relation

\begin{equation}
I_{c}(Q,\omega)=\left [\frac{\hbar \omega / KT }{1-e^{-\hbar
\omega / KT }}\right ]^{-1}I_{q}(Q,\omega) \label{classica}
\end{equation}

\noindent where the suffix c and q indicates the classical and
quantum quantities, respectively.

The good agreement between the simulated and the experimental
spectra testifies that the adopted potential represents well the
lithium system \footnote{The experimental resolutions of the
reported spectra are slightly different for the experiment and
simulation, in particular: $\delta \omega_{exp} =4.5$ ps$^{-1}$
and $\delta Q_{exp}=0.35$ nm$^{-1}$ FWHM, while $\delta
\omega_{sim} =1.3$ ps$^{-1}$ and $\delta Q_{sim}=0.2$ nm$^{-1}$
FWHM. The energy broadening of the spectra for the considered
$Q$-value is, however, always larger than any resolution
effects.}.

\begin{figure}[ftb]
\centering
\includegraphics[width=.53\textwidth]{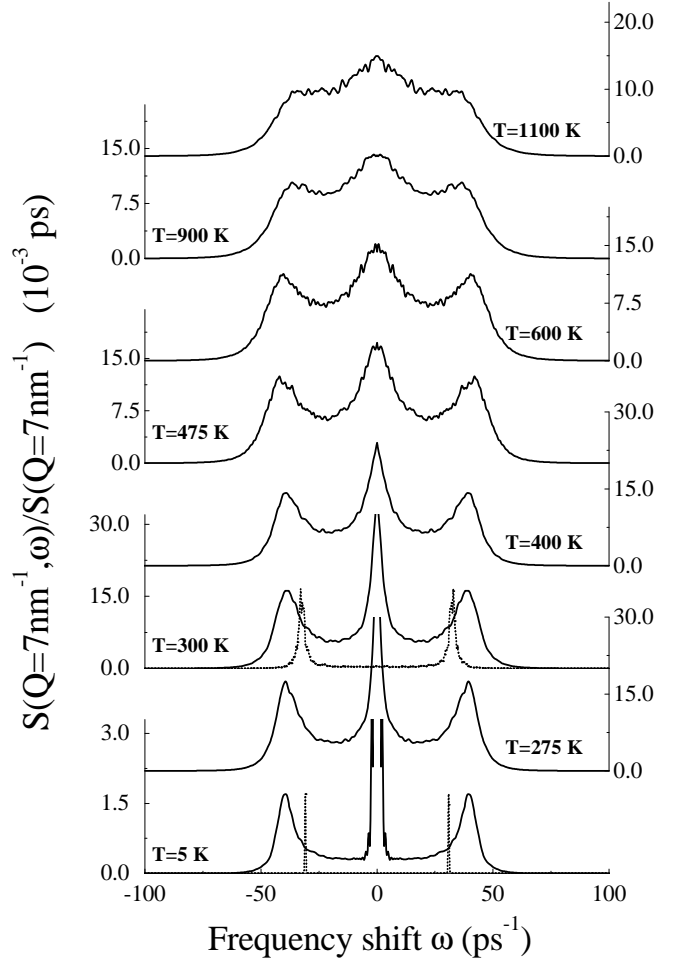}]
\vspace{-.2cm} \caption{ Dynamic structure factor of lithium at $Q=7$nm$^{-1}$
at different temperatures and phases:  liquid, supercooled and glass (full
line), crystal (oriented along (1 0 0), dashed lines). The energy resolutions
(FWHM) is $\delta E_{sim}=1.3$ ps$^{-1}$ (in the crystal case it has been
enhanced by a factor three due to the narrower lineshape) and in all cases it
is negligible with respect to the intrinsic spectral width.} \label{panel}
\end{figure}

In Fig.~\ref{panel} we report the temperature dependence of the
dynamical structure factor for the same selected value of the
momentum transfer, $Q=7.0$ nm$^{-1}$.

The temperature dependence of the spectra reported in
Fig.~\ref{panel} shows two apparently opposite behaviors; on one
side the central line -or Mountain peak in the terminology of the
glass transition phenomenology- gets narrower with decreasing
temperature and its width -related to the inverse of the
relaxation time $\tau$- becomes negligible at the
glass-transition, thus indicating the structural arrest (i.~e. a
divergent relaxation time). On the contrary the Brillouin
component shows a much reduced temperature dependence, and the
excitation maintains a significant linewidth even at very low
temperature and no remarkable behavior is observed around $T_c$.

It is worth to remember that the width of the Brillouin peaks is
a non-monotonic function of $T$, they attain a maximum in the
temperature region where $\Omega\tau\approx 1$, and become
negligible when $\Omega\tau >> 1$ (here $\Omega$ is the frequency
of the excitation, i.~.e. approximately the Brillouin peak
position). The origin of the non vanishing width of the Brillouin
peak has to be searched in a phenomenon different from the
relexation process responsible for the structural arrest
($\alpha$-process).

To investigate the effect of a possible origin (the anharmonicity)
of the observed Brillouin linewidth, we report the spectra
obtained by MD at 5 and 300 K in the lithium crystal, produced by
thermalizing at finite temperature the configuration obtained
starting from an ordered (BCC) lattice.

As clearly shown in the figure, and as well known from the experiments, the
phonon linewidth in the crystal is completely negligible compared to the
disordered phases, indicating that the residual linewidth cannot be attributed
to anharmonicity, which, in turn, is expected to be of the same order of
magnitude in the ordered and disordered phases.

\begin{figure}[ftb]
\centering
\includegraphics[width=.54\textwidth]{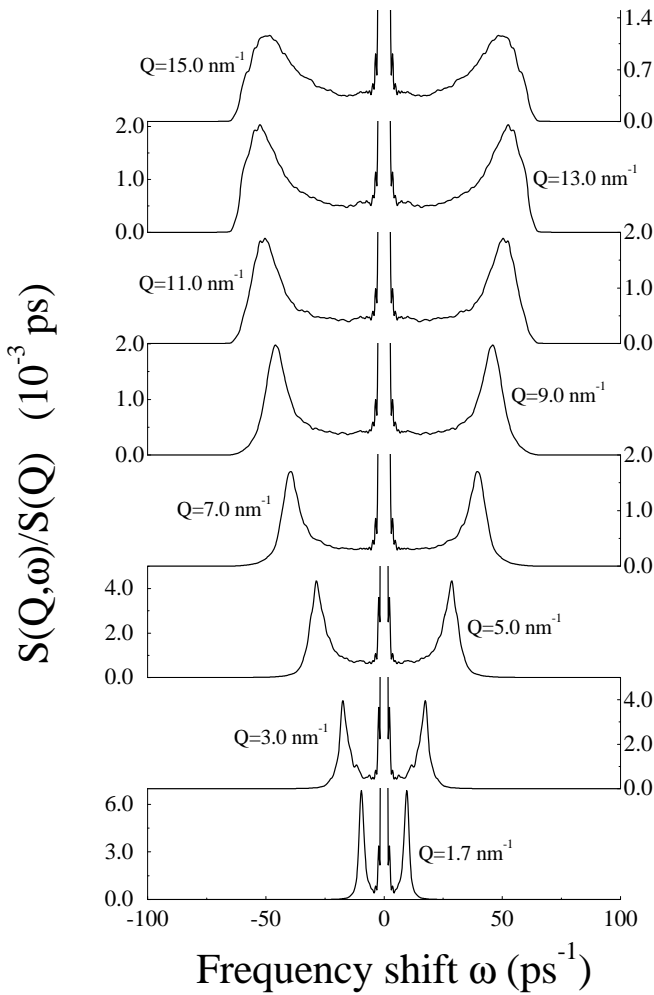}]
\vspace{-.2cm} \caption{Dynamic structure factor of a lithium
simulated glass at T=5 K.} \label{panelQ}
\end{figure}

Having excluded the anharmonic origin of the Brillouin linewidth, a
possibility lies in the presence of a further relaxation process, that in the
system parallels the $\alpha$-process. In order to investigate on such a
possibility, an appropriate way can be to study the dynamic structure factor
in the glassy phase, where any relaxation effect due to the $\alpha$ process is
frozen. The dynamic structure factor of a lithium glass, obtained by
"instantaneous" quench from above the melting point, is reported in
Fig.\ref{panelQ}.

\begin{figure}[ftb]
\hspace{-.7cm}
\includegraphics[width=.48\textwidth]{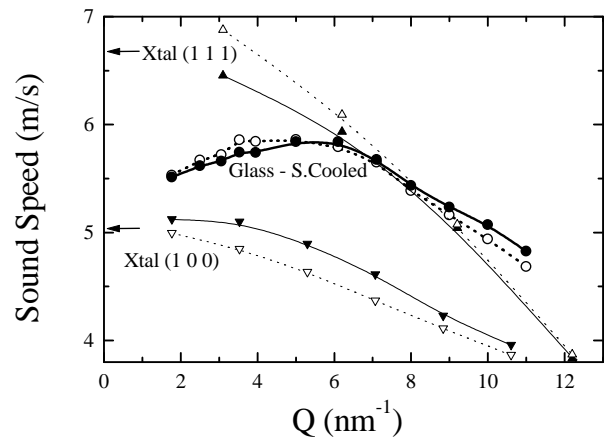}
\vspace{-4.cm} \caption{ Speed of sound in different
thermodynamic phases. Dots:glass/supercooled, downtriangle:
crystal along (1 0 0), uptriangle: crystal along (1 1 1). Full
and open symbols are relative to $T=5$ K and $T=300$ K,
respectively. Arrows indicates acoustic data at $T=70$ K. Lines
are guideline for the eyes only.} \label{crist}
\end{figure}

As a demonstration of the presence of such an additional relaxation process, we
report the observation of a positive dispersion of the sound velocity below
the melting temperature. This effect, also expected to take place at
frequencies such that $\omega\tau \approx 1$, is shown in Fig.~\ref{crist},
where we report the quantity $c_l(Q)=\omega_l(Q)/Q$ with $\omega_l$ the
maximum of the current correlation spectra

\begin{equation}
J(Q,\omega)=\omega^2/Q^2S(Q,\omega).
\end{equation}

The existence of a positive dispersion of sound, reproduced by
the simulations of the glass and cold liquid phases (it is
missing in the crystal), supports the hypothesis of the existence
of a second relaxation process.

The qualitative analysis of Fig.~\ref{panel} {\it per se}
indicates the presence of at least two relaxation processes
governing the high frequency collective dynamics of the
considered system: one, strongly temperature {\it dependent},
which shows up mostly in the width of the central line, and the
other, strongly temperature {\it independent}, which gives rise
to the width of the inelastic feature. These simulations confirm,
therefore, in a much wider temperature range the conclusions
provided by the analysis of the IXS data on lithium \cite{scop},
and extend into the supercooled and glassy phase regions the
theoretical predictions of a dynamics ruled by two different
time-scales. Particularly interesting is the persistence of the
second (fast) relaxation process, which is essentially identical
in the liquid and in the glass. This result is somewhat in
contradiction with the current belief: in fact, in the liquid the
fast dynamical processes has been so far associated to
uncorrelated binary collisions, while in the solid the atomic
motion is dominated by long range correlations. A quantitative
description of this relaxation process may contribute to shed
light on this point.

\begin{figure}[ftb]
\centering
\includegraphics[width=.4\textwidth]{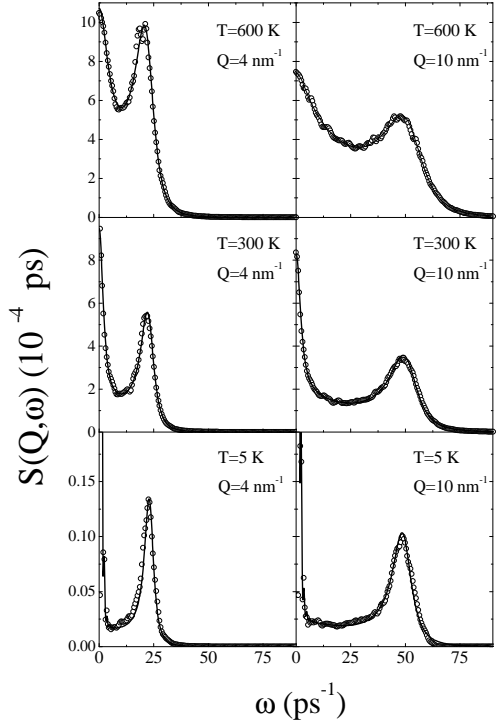}
\caption{ Seletecd fitting examples at different $Q-T$ values.
Open dots: molecular dynamics, full lines: fits. $S(Q,0)$ values
at $T=5$ K are $2.1 \times 10^{-3}$ and $4.9 \times 10^{-3}$,
respectively. } \label{fitting}
\end{figure}

We now quantitatively represent the dynamic structure factor to
asses the evolution of the  relaxation processes from the liquid
down to the supercooled and glassy state. To this purpose, we
analyzed our data adopting a model based on the memory function
approach, introduced to represent the MD data on alkali metals
\cite{mori,lev}, and recently utilized to interpret the
experimental data of liquid lithium \cite{PRL}, aluminum
\cite{PREal} and sodium \cite{PREna}. Within this framework, the
evolution of the density autocorrelation function is described by
a memory function trough the Langevin equation:

\begin{eqnarray*}
\stackrel{..}{\phi }(Q,t)+\omega _0^2(Q)\phi (Q,t)+\mbox{$\int_0^t$}M(Q,t-
t^{\prime })%
\stackrel{.}{\phi }(Q,t^{\prime })dt^{\prime }=0  && \label{langevin}
\end{eqnarray*}
where

\begin{equation}
\omega _0^2(Q)=[k_BT/mS(Q)]Q^2
\end{equation}

By Fourier transformation, the dynamic structure factor is given
by:

\begin{equation}
\frac{S(Q,\omega )}{S(Q)}=\frac{\pi^{-1}\omega _0^2(Q)M^{\prime }(Q,\omega )}{%
\left[ \omega ^2-\omega _0^2+\omega M^{\prime \prime }(Q,\omega )\right]
^2+\left[ \omega M^{\prime }(Q,\omega )\right] ^2}  \label{sqw}
\end{equation}

The complex memory $M(Q,t)$ contains all the interaction details and decays
over the characteristic time-scales of the system. The dynamics of a simple
liquid can be satisfactorily described by the two time exponentials ansatz
\cite{lev}:

\begin{equation}
M_L(Q,t)=\Delta _\alpha^2(Q)e^{-t/\tau _\alpha(Q)}
+\Delta _\mu^2(Q)e^{-t/\tau _\mu(Q)}  \label{somme}
\end{equation}

where the indexes $\alpha$ and $\mu$ indicate the usual $\alpha$-
(or structural-) and the fast- (or microscopic) process,
respectively. We used Eq.~(\ref{sqw}) as a model function. Its
convolution with the  simulation resolution $R(\omega)$ has been
utilized as a fitting function to the MD spectra:

\begin{equation}
F(Q,\omega)=\int S(Q,\omega' )R(\omega- \omega')d\omega '
\label{fitfunction}
\end{equation}

The results of such procedure are shown in Fig.~\ref{fitting},
where we report the calculated dynamic structure factors together
with their best fitted line-shapes at selected momenta and
temperatures.

\begin{figure}[ftb]
\centering
\vspace{-.5cm}
\hspace{-1.5cm}
\includegraphics[width=.44\textwidth]{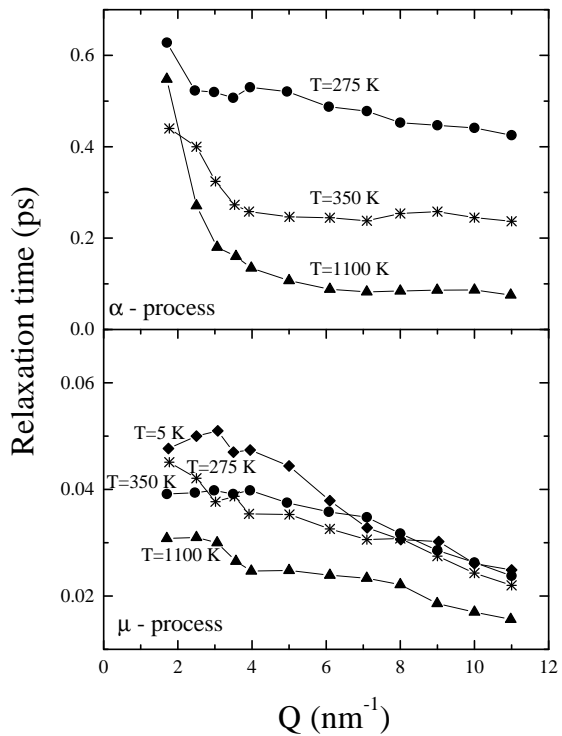}
\vspace{-.2cm} \caption{ Q-Dependence of the relaxation times at
some selected temperatures. The slight increase of the $\alpha$
process at low $Q$ is an artifact due to finite resolution
effects (see text). } \label{timesq}
\end{figure}

Among the fitting parameters, the relaxation times of the two
processes are reported in Fig.~\ref{timesq} as function of the
exchanged wavevector and at selected temperatures. At all
considered temperatures, the $\alpha$-process shows an almost
flat $Q$ dependence, with an apparent increase below $Q \approx
4$ nm$^{-1}$, an artifact that occurs when the relaxation time is
comparable or higher than the inverse resolution, already observed
in the analysis of the IXS spectra \cite{scop}. The microscopic
process shows, instead, an effective slightly decreasing $Q$
trend. The value of the structural relaxation time at large $Q$
shows a sharp increase in the supercooled region and upon further
cooling. On the contrary, the temperature behavior of the fast
microscopic, $\mu$, process is very different. In fact, it gives
a basically $T$-independent relaxation time in any of the
considered phases. This figure provides a quantitative account of
the qualitative analysis that was derived from the results
reported in Fig.~1.

\begin{figure}[ftb]
\centering
\includegraphics[width=.5\textwidth]{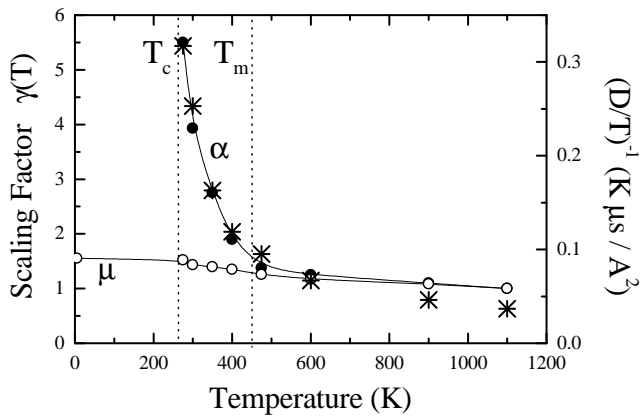}
\vspace{-6.8cm} \caption{ T-Dependence of the relaxation times (circles,left
axis). In the same plot we also report the behavior of the diffusion
coefficient (stars,right axis). } \label{timest}
\end{figure}

To be more precise in the assessment of the temperature dependence of the
relaxation times, once observed that the $Q-$ dependence is similar for each
temperature, we defined a $Q$-independent temperature coefficient,
$\gamma(T)$, through the relation $\tau (Q,T)=\gamma (T) \tau(Q,T_o)$. As
arbitrary reference temperature we selected $T_0=1100$ K, and we determined
$\gamma(T)$ as the scaling factor that minimizes the mean squared differences
between each set of data and the set at $T_o$.

The result of such a procedure, for both the $\alpha$ and $\mu$
processes, is reported in Fig. \ref{timest}. As clearly seen, the
relaxation time of the structural process increases about six
times between $T_o$ and the lowest available temperature ($T=275$
K). In the same plot we also report the quantity $T/D$, being $D$
the diffusion coefficient determined in the simulations. As
expected, this quantity follows a behavior similar to that of the
relaxation time of the $\alpha$-process, thus confirming the
validity of its identification with the structural relaxation
process. On the contrary, the microscopic relaxation time
exhibits a small variation in the whole temperature range with a
total change of only 50\% of the value at $T_o$, and without any
noteworthy behavior around $T_c$. This demonstrates that the fast
relaxation process is the same from the hot liquid down to the
glass.

\section{DISCUSSION AND CONCLUSIONS}

In the present paper, we have shown how the dynamics of a simple monatomic
system can be described in terms of two viscous relaxation processes in a
range of temperature covering all disordered phases: the hot liquid, the
supercooled liquid and the glass. Few comments concerning the physical origin
of these two processes are in order. By its strong temperature dependence and
by the close relation with the temperature dependence of the diffusion
coefficient, the slower relaxation process is unambiguously identified as the
$\alpha$ process, i.~e. the process that drives the glass transition in those
systems capable to sustain strong under-cooling. More interesting are the
indications obtained from the analysis of the faster process. The comparison
between glass and crystal, the equivalence of the fast relaxation process in
the liquid and in the glass, and the consistency between the simulation and
experimental results in the liquid, are all indications that the origin of this
process must be searched into something that is common between liquid and
glass, but not existent in the crystal. On the considered fast time scale,
which corresponds to values of fractions of ps, one can consider all diffusion
processes in the liquid frozen, and on this snap-shot time-scale one can
consider that the liquid and the glass are not only topologically, but also
dynamically equivalent. Along this line, one is then driven to conclude that
-similarly to what has been found in other model glasses \cite{har}- also in
the liquid this fast relaxation process is intimately related to the
topologically disorder, while anharmonicity and/or dynamical effects play only
a secondary role. The way in which the disorder can produce the phenomenology
of a "relaxation process" -also in the harmonic limit- is as follows. On a
general ground, a relaxation process can be pictured as the macroscopic
manifestation of microscopic phenomena associated with the existence of
channels by which the energy stored in a given "mode" relaxes towards other
degrees of freedom. The spectrum of the density fluctuations, $S(Q,\omega)$,
through the fluctuation-dissipation relation, reflects the time evolution of
the energy initially stored ($t$=$t_o$) in a plane wave of wavelength
$2\pi/Q$. As the plane wave is not an eigenstate of the disordered system, at
$t\!>\!t_o$ there will be a transfer of amplitude from this plane wave towards
other plane waves with different $Q$ values. This process is controlled by the
difference between the considered plane wave and the true normal modes of the
topologically disordered glassy structure. This energy flow takes place on a
specific time scale $\tau$, and gives rise to the observed relaxation process
phenomenology. The depicted scenario is in {\it contrast} with a description
of the fast relaxation process in terms of (more or less correlated) binary
collisions. Indeed, according to the usual description, the collision between
two atoms -at a given position in space $\bar r^*$ and at a given time $t^*$-
are not correlated with the dynamics of the atoms far away from $\bar r^*$,
and with the dynamics of the colliding atoms at time larger than
$t^*+\tau_\mu$. In this view $\tau_\mu$ acts a the correlation time of the
local dynamics. On the contrary, according to the present description, the
dynamic of the whole system is strongly correlated -and described in term of
normal modes of vibration- up to a time of the order of $\tau_\alpha$, and
$\tau_\mu$ indicates the decorrelation time for the specific plane waves, not
for the dynamics.

The common origin of the fast process in the liquid and in the glass, and its
association to the disorder, poses against its interpretation in terms of
kinetic processes such as ``binary collisions''. In this sense, the idea that
all the binary processes are contained in the short time dynamics ($THz$
region), while the correlated motion is fully described by the $\alpha$
process has to be revised.

\section{ACKNOWLEDGEMENTS}

The authors gratefully acknowledge U. Balucani for countless discussions on
the origin of the microscopic relaxation: the realization of this paper is also
due to his severe and constructive criticism. R. Di Leonardo, D. Fioretto, F.
Sciortino and P. Tartaglia are also acknowledged. T.S. is finally grateful to
W. Goetze for the encouragement that originated from the time spent discussing
on the subject of this paper.

\end{document}